INFLUENCE OF THE DUFOUR EFFECT ON STRIATIONS FORMATION IN RADIO-FREQUENCY DISCHARGES

Dmitry Levko[1,a] and Laxminarayan L. Raja[2]

[1]Esgee Technologies Inc., Austin, Texas 78746, USA

[2]Department of Aerospace Engineering and Engineering Mechanics, The University of Texas at Austin, Austin, Texas 78712, USA

[a] e-mail address: dima.levko@gmail.com

In recent years, interest in striation phenomena in radio-frequency (rf) discharges has risen due to the availability of new experimental data and implementation of new computational models. Depending on the conditions, different mechanisms of discharge striations are realized. These are the ionization instability, the instability due to the electron attachment to electronegative gases or the instability due to thermoelectric transport. Although the first two mechanisms were modeled quite extensively in recent years, the understanding of the influence of Dufour effect originating from plasma density gradients on stability of radio-frequency discharges in long tubes remains poor. In this paper, the influence of this mechanism on the longitudinal striations of radio-frequency discharge is presented using a one-dimensional model of argon discharge driven with rf excitation under intermediate pressure conditions of 0.5 Torr. It is found that striation formation is sensitive to the value of the thermoelectric heat transport coefficient in the low electron temperature range. The critical value of this coefficient necessary for the instability onset is derived using the linear stability analysis.

## I.    Introduction

Striations in glow discharges is one of the most fascinating phenomena in discharge physics.[1,2] They have been observed in rare and molecular gases, at high and low gas pressures, in magnetized and unmagnetized plasmas of direct current (dc) and radio frequency (rf) discharges (see Refs. 3,4,5,6,7 and references therein). Due to such disparity of physical scales, different physical mechanisms govern the formation of striations.[2] These can be kinetic effects, non-linear dependence of the ionization frequency on the electron density, electron attachment to electronegative gases, thermoelectric heat transfer *etc*.

Striations of positive column of glow discharges was studied in numerous publications using fluid, kinetic and hybrid approaches.[8,9,10,11,12] Less attention was devoted to the computational analysis of rf discharge striations. The striations of electronegative gas $CF_4$ was studied in Ref. 3. It was found that the striations are generated due to the resonance between the rf frequency and the eigenfrequency of the ion-ion plasma. This frequency establishes a modulation of the plasma parameters and the electron transport.

Pattern formation in low-pressure inductively coupled plasma was studied in Ref. 6. Linear stability analysis has shown that striations appear at low and intermediate pressures due to electron energy transport. This mechanism was used in Ref. 13 to explain the formation of striations in rf discharges of argon and nitrogen forming perpendicular to the applied electric field. The authors used a two-dimensional fluid model coupled with the Poisson's equation. In this model, the discharge instability was enforced by the random perturbation of the electron density by 1%. The longitudinal striations of rf discharge due to the ionization instability was modeled in Ref. 7 using the one-dimensional fluid drift-diffusion model. The nonlinear dependence of the ionization frequency on the electron density was used in these studies. Standing striations were observed for a wide range of rf frequencies and gas pressures. The electron kinetics for similar conditions was studied in Ref. 12.

Thermocurrent instability of direct current discharge in various gases was studied in Refs. 14,15,16. It was shown that this instability arises due to the competition between the diffusion and thermodiffusion electron fluxes. The electron density and temperature gradients are directed in opposite directions. Then, if the thermodiffusion flux exceeds the diffusion flux, any plasma perturbation will grow with time resulting in thermocurrent instability.

To our knowledge, the self-consistent modeling of striations formation in rf discharges in long tubes was attempted only in Refs. 7 and 12. In Ref. 7, the mechanism of discharge striations was prescribed by introducing the special form of the ionization frequency, while the mechanism of discharge striations in Ref. 12 was not clear. In the present paper, the influence of Dufour effect originating from plasma density gradients within the plasma on striations of rf discharge at the gas pressure of 0.5 Torr is studied by the drift-diffusion fluid model. At such a pressure, the fluid approach to the description of plasma is justified due to the frequent electron-neutral collisions. Also, it is important to note that for the conditions of our studies, the angular rf frequency is much higher than the electron energy relaxation frequency (see discussion in Section III.A). Therefore, the plasma density remains constant while the electric field oscillates with the rf frequency.[17]

The paper is organized as follows. Section II discusses the fluid model used in the present studies. Section III presents the analysis of the influence of the thermoelectric heat transfer on the discharge stratification. Section IV summarizes the results of the present studies.

## II.    Numerical model

In the present studies, the one-dimensional fluid model for two-component plasma consisting of electrons and argon ions (Ar⁺) is used. The dominant background neutral (Ar) "gas" density was kept constant and homogeneous throughout the entire domain. Also, the generation of the electronically excited argon species and consequent stepwise ionization were neglected in the present model. These processes may be important for the conditions of our studies, but do not affect the main objective of our study in elucidating the role of the thermoelectric heat transfer coefficient on striation formation.[17]

The model solves fluid equations for the species continuity with the drift-diffusion approximation:

$$\frac{\partial n_{e,i}}{\partial t} + \frac{\partial \Gamma_{e,i}}{\partial x} = k_{ion} n_e n_g - \frac{n_{e,i}}{\tau(x)}. \tag{1}$$

Here, $n_{e,i}$ and $\Gamma_{e,i}$ are the number density and number flux of electrons and ions, respectively, $k_{ion}$ is the ionization rate coefficient tabulated as the function of the electron temperature $T_e$, $n_e$ is the electron number density and $n_g$ is the background gas density. The last term on the right-hand side of Eq. (1) simulates the charged species ambipolar diffusion in the radial direction. This diffusion time is defined as $\tau(x) = \frac{\Lambda^2}{D_a(x)}$, where $\Lambda = R_{tube}/2.4$ is an effective transverse discharge dimension ($R_{tube}$ is the discharge tube radius) and $D_a$ is the ambipolar diffusion coefficient. The density of the background gas argon, $n_g$, is assumed to be homogeneous and constant.

The species flux terms for both charged and neutral species are calculated using the drift-diffusion approximation:

$$\Gamma_{e,i} = -s_{e,i} \mu_{e,i} n_{e,i} \frac{d\varphi}{dx} - \frac{\partial(D_{e,i} n_{e,i})}{\partial x}. \tag{2}$$

Here, $\mu_{e,i}$ and $D_{e,i}$ are the species mobility and diffusion coefficients, $s_i = 1$ and $s_e = -1$. Electron transport properties are computed as the functions of electron temperature using the Boltzmann equation solver Bolsig+,[18] which solves the Boltzmann equation in the two-term approximation. The Ar$^+$ transport properties were obtained from Ref. [19]. The electrostatic potential is obtained by solving the Poisson's equation:

$$\frac{d^2\varphi}{dx^2} = -\frac{q_e}{\varepsilon_0}(n_i - n_e),$$ (3)

where $q_e$ is the elementary charge. The electron energy density $n_\varepsilon$ is defined as

$$n_\varepsilon = \frac{3}{2}n_e q_e T_e,$$ (4)

where $T_e$ is in the units of eV. It is calculated by solving the electron energy conservation equation:[13]

$$\frac{\partial n_\varepsilon}{\partial t} + \frac{\partial \Gamma_\varepsilon}{\partial x} = S_\varepsilon.$$ (5)

Here, $\Gamma_\varepsilon$ is the flux of electron energy and $S_\varepsilon$ is the source term. They are defined by[18]

$$\Gamma_\varepsilon = -\mu_\varepsilon E n_\varepsilon - \frac{\partial(D_\varepsilon n_\varepsilon)}{\partial x},$$ (6)

$$S_\varepsilon = -q_e \Gamma_e E - q_e(\varepsilon_{ion} k_{ion} + \varepsilon_{ex} k_{ex})n_e n_g - q_e n_e \frac{2m_e}{m_i}(T_e - T_g)\nu_m,$$ (7)

In Eq. (7), $E$ is the electric field obtained as $E = -d\varphi/dx$, $\varepsilon_{ion}$ and $\varepsilon_{ex}$ are the thresholds for Ar ionization and excitation to the first electronic level (in eV), $k_{ex}$ is the rate coefficient of the excitation reaction of the first electronic level of Ar, $\nu_m$ is the electron-argon momentum transfer collision frequency, $m_e$ is the electron mass, $m_i$ is the argon mass, and $T_g$ is the background gas temperature. Also, $\mu_\varepsilon$ and $D_\varepsilon$ are, respectively, the electron energy mobility and diffusion coefficient calculated by Bolsig+.

The first term on the right-hand side of Eq. (7) describes the electron heating by the electric field (further ohmic heating), the second term describes the contribution of inelastic collisions, and the last term describes the contribution of elastic collisions. For further discussion, it is also convenient to introduce the thermoelectric energy transport coefficient defined as

$$\chi_e = \left(\frac{\beta_e}{\mu_e} - \frac{G_e}{D_e}\right) D_e. \tag{8}$$

Here, $G_e$ and $\beta_e$ are, respectively, the coefficients of heat diffusion and coefficient of thermoelectric transport for electrons. According to Ref. [18], they are defined by solving the Boltzmann equation as $\beta_e = \bar{\varepsilon}\mu_\varepsilon$ and $G_e = \bar{\varepsilon}D_\varepsilon$, where $\bar{\varepsilon}$ is the average electron energy.

Here, it is important to note that Eq. (5) in which $\mu_\varepsilon$ and $D_\varepsilon$ are obtained from the EEDF is consistent with the two-term approximation of the Boltzmann equation. Another form was used in Ref. [13] which explicitly includes the thermoelectric flux defined as the coefficient (8) multiplied by the electron density gradient. This form is inconsistent with the two-term approximation.[18]

Kinetic Maxwellian flux condition combined with secondary electron emission flux is used to specify the boundary flux of the electrons to electrode surfaces:

$$\vec{\Gamma}_e \hat{n} = \frac{1}{4} n_e \sqrt{\frac{8 q_e T_e}{\pi m_e}}, \tag{9}$$

Here, $\hat{n}$ is the unit normal vector pointing towards the electrode surface from the plasma. In the present studies the secondary electron emission from the walls was neglected.

Mobility limited flux condition is imposed at the walls for ions using

$$\vec{\Gamma}_{ion} \hat{n} = \frac{1}{4} n_i \sqrt{\frac{8 q_e T_g}{\pi m_i}} + n_i \max(0, -\mu_i \hat{n}_s \frac{\partial \varphi}{\partial x}). \tag{10}$$

For the electron energy equation, the energy flux incident at the walls is given by

$$Q_\varepsilon^w = \frac{5}{2} q_e T_e \Gamma_e^w. \tag{11}$$

Here, $\Gamma_e^w$ is the electron number flux at the walls given by Eq. (10).

Equations (1) and (5) were discretized using the Scharfetter-Gummel spatial discretization scheme.[20,21] A uniform mesh with 2000 grid cells is used for the simulations and the time step of $10^{-10}$ s was used for all cases considered.

In the present studies, the interelectrode distance was set equal to 30 cm. The background gas pressure was 0.5 Torr and the background gas temperature was 300 K. The right boundary was kept grounded while the rf potential was applied to the left boundary, $U_{left}(t) = U_0 \sin(2\pi f t)$ with $U_0 = $ -400 V and $f = 13.56$ MHz. The tube radius was assumed equal to $R_{tube} = 1.1$ cm. The secondary electron emission from the boundaries was neglected in the present studies.

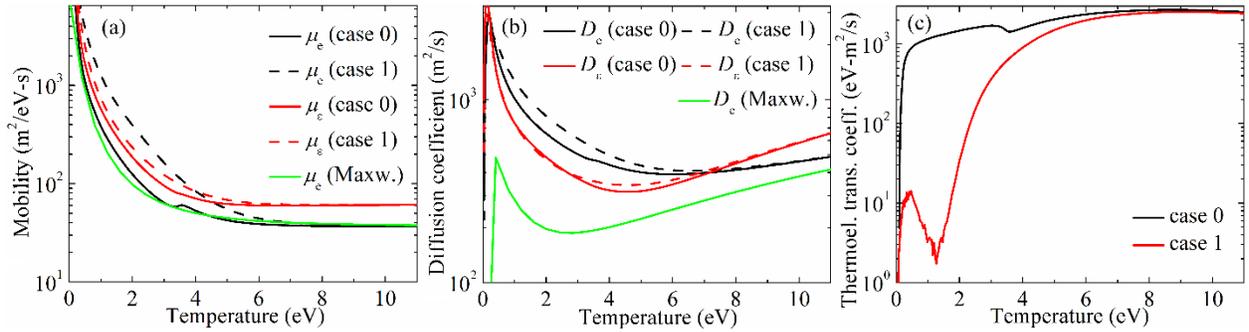

Figure 1. (a) Electron mobility, (b) electron diffusion coefficient and (c) thermoelectric energy transport coefficient obtained for three different assumptions on the electron energy distribution function.

Figure 1 shows three sets of the electron transport coefficients used in the present studies. These coefficients were obtained by using the solver Bolsig+.[18] Three cases were considered. First, it was assumed that the electron energy distribution function (EEDF) is Maxwellian. In this case, one obtains $\mu_\varepsilon = \frac{5}{3}\mu_e$ and $D_\varepsilon = \frac{5}{3}D_e$,[18] and, consequently, $\chi_e = 0$. In case 0, the non-equilibrium conditions were considered without accounting for the electron-electron Coulomb collisions. In case 1, to understand the influence of the electron-electron collisions on the Dufour effect the plasma ionization degree of $10^{-4}$ was set in the solver Bolsig+. Here, it is important to note the rate coefficients of the electron-neutral reactions considered in our studies also depended on the EEDF's choice.

### III.     Results and discussion

### A.     Maxwellian vs. non-Maxwellian electron energy distribution function

In this subsection, we compare case 0, i.e. the case of non-Maxwellian EEDF without accounting for the electron-electron collisions, with the case of the Maxwellian EEDF. In the steady state, the electron energy flux defined by Eq. (6) may be simplified.[6,9] Let us consider the quasi-neutral plasma of positive column. Then, the transport of both electrons and ions is described by the ambipolar flux defined as

$$\Gamma_e = \Gamma_i \approx -\frac{\mu_i}{\mu_e}\frac{\partial(n_e D_e)}{\partial x},$$  (12)

while the ambipolar electric field within the plasma is

$$E \approx -\frac{1}{n_e \mu_e}\frac{\partial(n_e D_e)}{\partial x}.$$  (13)

Then, the electron energy flux becomes:

$$\Gamma_\varepsilon = \chi_e \frac{\partial n_e}{\partial x} + \kappa_e \frac{\partial T_e}{\partial x}.$$  (14)

Here, the thermoelectric energy transport coefficient is defined by Eq. (8) and the thermal conductivity is defined as

$$\kappa_e = \left(\frac{\beta_e}{\mu_e}\frac{dD_e}{dT_e} - \frac{dG_e}{dT_e}\right)n_e.$$  (15)

Equation (14) shows that in the case of non-equilibrium EEDF the flux consists of two competing components proportional to the electron density and temperature gradients. The term with $\chi_e$ leads to the flux similar to the Dufour effect discussed, for instance, in Ref. 22. For the Maxwellian EEDF one obtains $\chi_e = 0$, while the thermal conductivity is defined as $\kappa_e = -\frac{5}{2}k_B n_e D_e$. Then, the energy flux for the Maxwellian EEDF is reduced to

$$\Gamma_\varepsilon = -\frac{5}{2}k_B n_e D_e \frac{\partial T_e}{\partial x},$$  (16)

i.e. it is defined by the electron temperature gradient only. Namely, it is directed opposite to $\frac{\partial T_e}{\partial x}$, i.e. leads to the energy flow from the regions of hotter plasma to the colder regions. In such plasmas, the Dufour effect cannot be obtained. In the case of non-Maxwellian EEDF, flux $\chi_e \frac{\partial n_e}{\partial x}$ may have an opposite direction and, as will be discussed in Section III.B, destabilize discharge.[16]

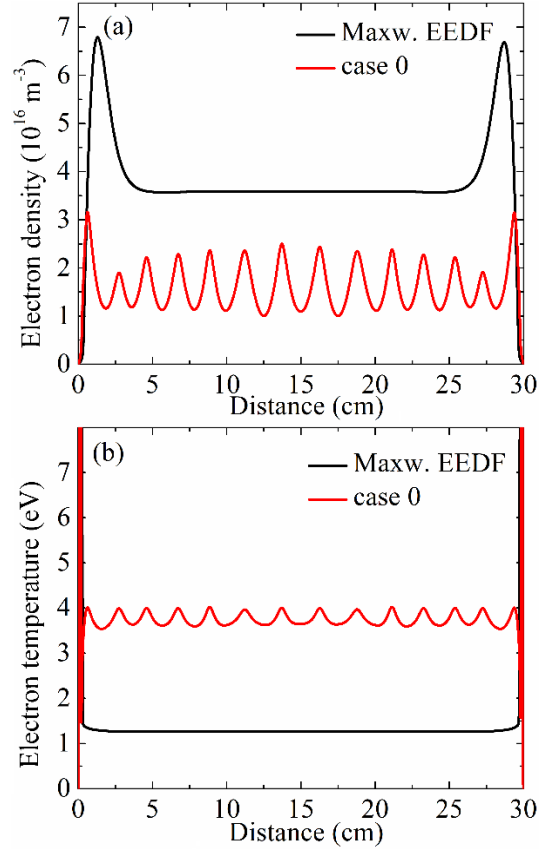

Figure 2. (a) Period-averaged electron density and (b) period-averaged electron temperature obtained with and without accounting for the thermoelectric effect.

The spatial distribution of the period-averaged plasma parameters for the Maxwellian and non-Maxwellian EEDFs is shown in Figure 2 and Figure 3, respectively. One can see that for $\chi_e = 0$ the discharge structure resembles the structure of dc glow discharge in long tubes.[23] Namely, there are the sheaths of non-neutral plasma in the vicinity of the electrodes, and the long positive column of quasi-neutral plasma. For the observation of such a structure, the last term on the right-hand side of Eq. (1) describing the plasma ambipolar radial diffusion to the side wall is crucial. Without accounting for this term, the plasma density reaches unrealistically large values, $\sim 10^{20}$-$10^{21}$ m$^{-3}$ due to the slow plasma diffusion from the discharge center to the left and right electrodes.

Unlike the dc discharge, rf discharge does not have the anode layer.[23] Here, both electrodes act as the anode each half of the rf period. As in the case of dc discharge, the sheaths are crucial for the discharge maintenance since the external electric field is shielded by quasi-neutral plasma. Therefore, the main potential drop is obtained near the electrodes [see Figure 3(a)]. Hence, the highest electron ohmic heating is also obtained within the sheaths [see Figure 3(b)], where the oscillating electric fields heat up the plasma electrons reaching the electrodes during the sheaths collapse. This results in the highest electron temperature [see Figure 2(b)] within the sheaths.

Figure 2(a) shows that the peaks of the plasma density are obtained near the electrodes. Here, the local maxima of the electrostatic potential are obtained (not shown here) which results in the reverse of the electric field in this region [see Figure 3(a)]. Such potential structure is formed because the electrons being accelerated within the sheaths dissipate their energy entering the quasi-neutral plasma where electric field is much smaller.

Since for $\chi_e = 0$ the plasma density is homogeneous within the positive column, the electron transport from the sheaths toward the discharge center is driven by the weak electric field present within the plasma. That is, the electron flux here is defined by the drift component [the first term on the right-hand side of Eq. (2)], while the diffusion component is zero due to zero electron density gradient. The same is true for electron energy transport since the gradient of the electron energy density is zero within the positive column. Since the electron-ion recombination is not considered in the present study, the electrons carried by the ambipolar electric field from the sheaths towards the discharge center are lost only through radial diffusion to the side wall.

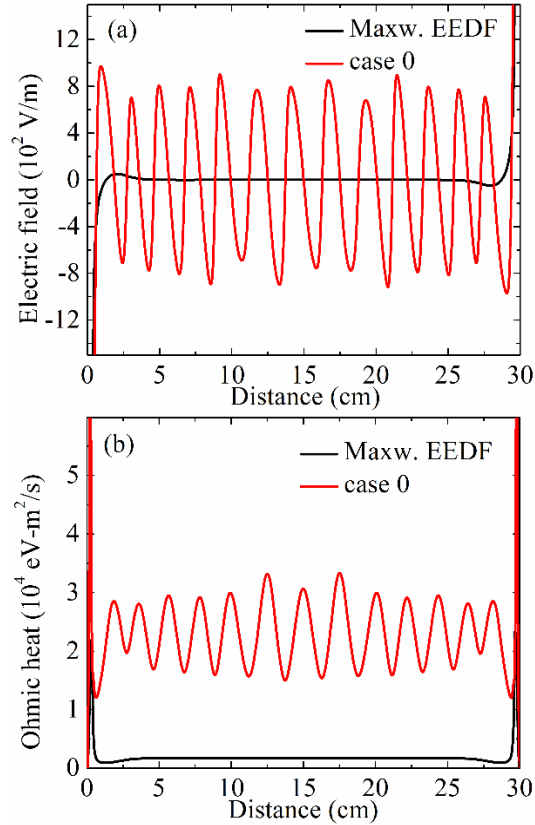

Figure 3. (a) Period-averaged electric field and (b) period-averaged ohmic heating of electrons obtained with and without accounting for thermoelectric effect.

The switching to the non-equilibrium EEDF changes the discharge structure drastically (see Figure 2). One can see the stratification of the positive column into 14 standing peaks in the electron density. Unlike in Ref. 7 where the generation of striations was enforced by the non-linear dependence of the ionization rate coefficient on the electron density, in the present studies they are initiated self-consistently. As will be discussed in more detail in Subsection III.B, they originate at the sheath edges and facilitate the electron heat transfer from the sheaths towards the discharge center. Here, it is also important to note that in the simulations presented in Ref. 13, the transverse striation formation was initiated by randomly perturbing the electron density, i.e. by introducing the numerical noise. In our simulations, the striations are formed self-consistently without the explicit introduction of any noise although there still can be some noise associated with the numerical discretization of plasma equations.

Figure 3(b) shows the period-averaged electron ohmic heating. One can see that for the stratified discharge, the dominant heating is still obtained within the plasma sheaths where the electric field is the highest. However, now there is also significant electron heating within the plasma bulk where the electric field ~400 V/m is obtained [Figure 3(a)]. One can conclude from Figure 3(b) that the total energy gain by the electrons within one striation is comparable to that obtained within the sheaths. This is due to both the electric field present within the striated plasma and the large electron density within one striation.

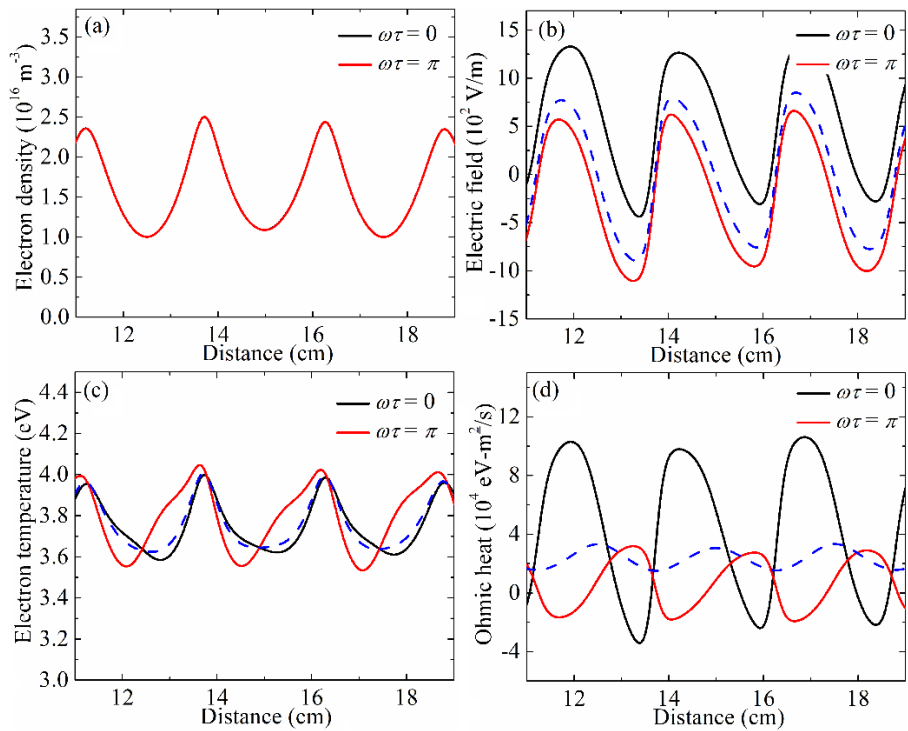

Figure 4. Spatial distributions of (a) the electron density, (b) electric field, (c) electron temperature and (d) power deposition from the electric field obtained near the center of the simulation domain at two different times of rf period. Dash lines show the period-averaged values.

Figure 4 shows the electron density, electric field, electron temperature and power deposition from the electric field to the electrons at two different times of the rf period. One can see that the electron density does not change with time, while the electron temperature exhibits modulations around its average value. Figure 4(b) shows the oscillations of electric field during one rf period which results in the oscillations of the electron ohmic heating term [Figure 4(d)].

Such dependence of the plasma parameters can be understood by comparing the typical frequencies with the rf frequency. For the conditions of our studies, the frequency of electron loss to the radial wall is estimated as $\nu_a = \frac{\mu_i T_e}{\Lambda^2} \sim 0.3 \times 10^6$ s$^{-1}$, where the ion mobility is $\mu_i \approx 1.6$ m$^2$/V-s. The collision frequency is estimated as $\nu_C = \delta \nu_m + \nu_{ion} + \nu_{ex}$, where for argon gas $\delta = \sqrt{2m_e/m_i} \sim 5.2 \times 10^{-3}$, $\nu_{ion}$ is the ionization frequency, and $\nu_{ex}$ is the electronic excitation frequency. For $T_e = 4$ eV [see Figure 2(b)], one has $\nu_m \approx 2.4 \times 10^9$ s$^{-1}$, $\nu_{ion} \approx 3.2 \times 10^4$ s$^{-1}$ and $\nu_{ex} \approx 1.4 \times 10^6$ s$^{-1}$. Thus, the collision frequency is $\nu_C \approx 1.3 \times 10^7$ s$^{-1}$, which is much larger than $\nu_a$. One can conclude that for the conditions of the present studies, discharge operates in the so-called high-frequency regime[24] since $\omega > \nu_C$ in which the plasma density is defined by the "effective" electric field and do not depend on its instantaneous values. This is seen in Figure 4(a).

The discharge striations change the mechanism of electron mass and heat transfer between the electrodes. Figure 5(a) shows the comparison between the period-averaged drift and diffusion components of the electron flux in Eq. (2) and Figure 5(b) shows the difference between their absolute values. One can see that unlike in the case of $\chi_e = 0$ where the electrons are carried by the small electric field presented within the plasma, in the case of non-Maxwellian EEDF the main mechanism of electron transport is their diffusion due to the density gradient within one striation. Note that the diffusion component does not oscillate since the electron density is independent of the oscillating electric field (see Figure 4). The drift component exhibits oscillations due to the modulations of the electric field like those shown in Figure 4(d) for the ohmic heating term.

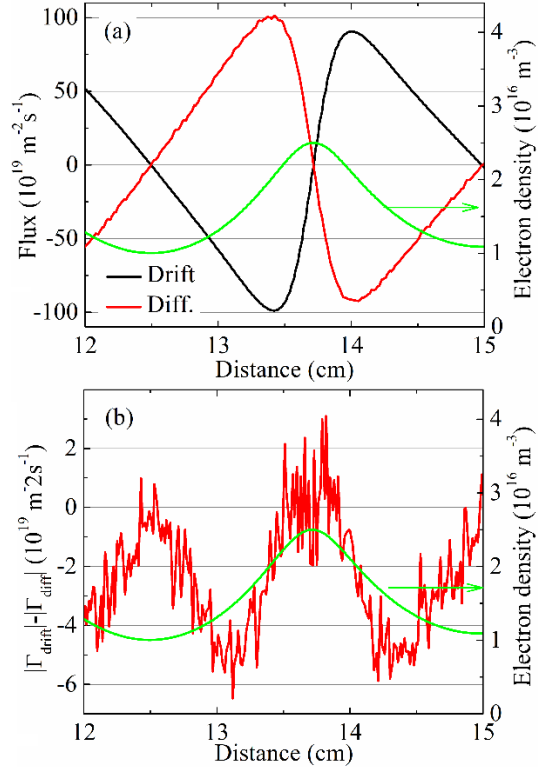

Figure 5. (a) Comparison between the period-averaged drift and diffusion components of the electron flux and (b) the difference between the absolute values of the drift and diffusion components of the electron flux for case 0 in the vicinity of one striation.

## B.    Influence of the electron-electron collisions on discharge structure

In order to clarify the nature of the obtained striations, this subsection analyzes the influence of the electron energy transport coefficient on the rf discharge structure. For this, two cases discussed in Section II were considered. In case 0 ($\chi_e = \chi_{e,0}$), the influence of the electron-electron collisions on the electron energy distribution function was ignored. In case 1 ($\chi_e = \chi_{e,1}$), these collisions were considered which resulted in a small value of the thermoelectric coefficient for $T_e < 3$ eV [see Figure 1(c)]. Figure 6 shows the comparison between the electron density and the electron temperature obtained in these two cases. One can see that there is no discharge stratification for $\chi_e = \chi_{e,1}$.

Equation (14) shows that the electron energy flux is defined by two competing components proportional to the electron density and temperature gradients. The comparison between these two components for two cases considered in this subsection is shown in Figure 7. The sum of these two terms is balanced by electron ohmic heating and the electron energy dissipation in inelastic collisions. The momentum transfer term is much smaller than other terms due to the small ratio between the electron and neutral masses and can be neglected in our analysis.

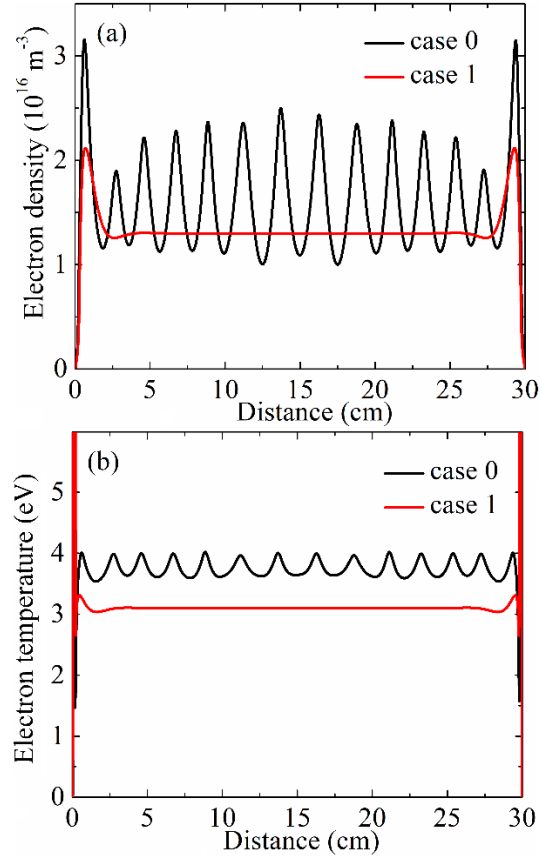

Figure 6. The comparison between (a) the electron density and (b) the electron temperature obtained for two values of the thermoelectric coefficient.

For the non-striated discharge, derivatives of both fluxes are non-zero only in the vicinity of the electrodes. In the plasma bulk both terms are almost zero due to zero electron density and temperature gradients. In this region, the electron ohmic heating is balanced by the energy dissipation in inelastic collisions.

Figure 7(a) shows that for $\chi_e = \chi_{e,0}$ both fluxes are directed in the opposite direction. The thermoelectric heat flux is in the direction of the electron density gradient and accumulates energy in the regions of higher $T_e$. The thermal flux $\kappa_e \frac{\partial T_e}{\partial x}$ is directed opposite to the electron temperature gradient and carries the energy from the striation. One can see from Figure 7(a) that $|\chi_e \frac{\partial n_e}{\partial x}| > |\kappa_e \frac{\partial T_e}{\partial x}|$. Thus, on average the thermoelectric flux dominates the electron energy transfer, which means that "hotter" regions of plasma are getting "hotter." It is important to note that the flux $\chi_e \frac{\partial n_e}{\partial x}$ does not change with time in the plasma bulk because the electron density is constant here. At the same time, the thermal flux follows the oscillations of the electron temperature gradient. Therefore, there are the times during one rf period when both fluxes add up which facilitates the energy transfer.

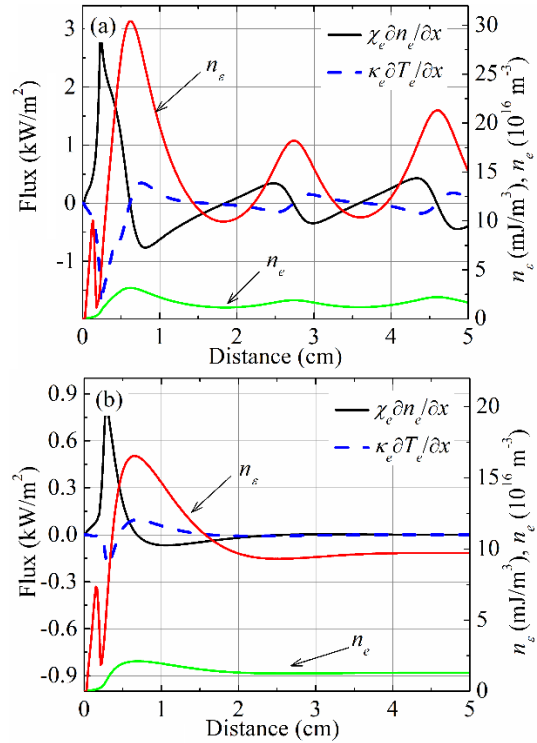

Figure 7. Comparison between the period-averaged fluxes obtained for (a) $\chi_{e,0}$ and (b) $\chi_{e,1}$.

Figure 7(b) shows that for $\chi_e = \chi_{e,1}$ both fluxes are non-zero only near the electrode, at $x < 2$ cm, where the gradients of the electron density and temperature are obtained. In the plasma bulk, both fluxes are zero and the electron energy is carried only by the electric field. As it follows from Figure 1(c), the thermoelectric coefficient for case 1 is ~1 order of magnitude smaller than for case 0. This means that the role of the Dufour effect decreases for increasing role of the electron-electron collisions on the EEDF formation.

One can conclude from Figure 7(a) that for $\chi_e = \chi_{e,0}$ the extrema of both fluxes are obtained within the sheaths. For instance, near the left electrode they are obtained at $x \approx 0.25$ cm. This distance agrees with the mean free path of electrons being accelerated within the sheaths and entering the quasi-neutral plasma bulk. For energetic electrons obtained within the sheaths, this is also the electron energy relaxation length. For these electrons, the dominant collision is the ionization one which is ~$2 \times 10^{-20}$ m$^2$. Then, for the gas pressure of 0.5 Torr the mean free path is estimated as $\lambda \approx 0.3$ cm.

After propagating the distance of 0.25 cm, hot electrons enter the plasma with a weak electric field within it and start dissipating their energy. This leads to the local decrease of the electron temperature [see Figure 6(b)] and the electron energy density [see Figure 7(a)]. Since the electron density at the sheath edge is still growing [see Figure 7(a)], the thermoelectric flux does not change its direction. Also, at the sheath edge the gradient of the electron density starts growing faster than within the sheath. Figure 1(c) shows that for $\chi_e = \chi_{e,0}$ the thermoelectric coefficient remains small in the temperature range $1 - 2$ eV. Thus, the thermoelectric flux obtains the local maxima at the sheath edge. This energy is carried by the thermoelectric flux from the sheath to the striations closest to the electrodes. Similar explanation is valid for the flux $\kappa_e \frac{\partial T_e}{\partial x}$ which is directed opposite to the electron temperature gradient.

### C. Linear analysis of the instability growth

For the linear analysis of the instability growth, let us consider the steady state quasi-neutral plasma of the positive column. Then, the ambipolar flux and electric field are defined, respectively, by Eqs. (12) and (13), and the electron energy flux is defined by Eq. (14). The solutions of plasma equations are looked in the form:

$$n_e \approx n_0 + \delta n_e e^{-i\Omega t + ikx}, \quad T_e \approx T_0 + \delta T_e e^{-i\Omega t + ikx}.$$

Here, $n_0$ and $T_0$ are, respectively, the unperturbed values of the electron density and temperature, $\Omega = \omega + i\gamma$ with $\omega$ is the angular frequency and $\gamma$ is the instability growth rate, and $k$ is the wave number.

Let us assume that the EEDF is close to the Maxwellian. Then, the electron thermal conductivity is $\kappa_e \approx -\frac{5}{2} k_B n_e D_e$. Also, let us assume the Einstein relation between the electron diffusion coefficient and mobility, i.e. $D_e = T_e \mu_e$. Then, using the observation that $n_0 \delta T_e \gg T_0 \delta n_e$[9] and neglecting the term $k^2 \mu_i$ compared to $\nu'_{ion}$, one obtains from the particle balance equation (1):

$$(-i\Omega + k^2 \mu_i T_0)\delta n_e - n_0 \nu'_{ion} \delta T_e = 0. \tag{17}$$

Here, $\nu'_{ion} = \frac{d(n_g k_{ion})}{dT_e}$ is the derivative of the ionization frequency at $T_e = T_0$. We also took into account the fact that in the steady state the local electron density in the positive column is defined by the balance between the ionization and wall losses, i.e. $n_g k_{ion}(T_0) = \frac{1}{\tau}$. By neglecting the elastic collision term in the electron energy balance equation, we find:

$$\left(-\frac{3}{2} i\Omega T_0 - k^2 \chi_0 + \varepsilon_{ion} \nu_0\right) \delta n_e + \left(-\frac{3}{2} i\Omega n_0 + 5k^2 \mu_e n_0 T_0 + \varepsilon_{ion} n_0 \nu'_i\right) \delta T_e = 0. \tag{18}$$

Thus, the system of equations (17) and (18) defines $\delta n_e$ and $\delta T_e$. The determinant of this system must be zero which results in the system of equations for the angular frequency and the instability growth rate.

The equation for the imaginary part gives $\omega = 0$ which is expected for the rf discharge where the standing striations are obtained (see Section III.A). The equation for the real part allows obtaining the following dispersion relation:

$$\gamma(k) \approx -k^2 \mu_i T_0 + \frac{(k^2 \chi_0 - \varepsilon_{ion} \nu_0) \nu'_{ion}}{5k^2 \mu_e T_0 + \varepsilon_{ion} \nu'_{ion}}. \tag{19}$$

Analogous relation was obtained in Ref. 9 for dc discharges. One can see that the instability growth rate does not depend on the electron density but is defined by the ionization and the electron heat and mass transfer. The maximum value of the wave number is:

$$k_{max}^2 = \frac{1}{5\mu_e T_0}\left[\left(\frac{\varepsilon_{ion} \nu'_{ion}(\chi_0 \nu'_{ion} + 5\mu_e T_0 \nu_0)}{\mu_i T_0}\right)^{\frac{1}{2}} - \varepsilon_{ion} \nu'_{ion}\right]. \tag{20}$$

By introducing $\alpha = \frac{\chi_0 \nu'_{ion}}{5\mu_e \mu_i T_0^2}$, $\beta = \frac{\varepsilon_{ion} \nu'_{ion}}{5\mu_e T_0}$ and $\delta = \frac{\nu_0}{\mu_i T_0}$, this equation can be written as

$$k_{max}^2 = (\alpha\beta + \beta\delta)^{\frac{1}{2}} - \beta. \tag{21}$$

Then, the maximum value of the instability growth rate is

$$\gamma_{max} = \mu_i T_0 \left(\alpha + \beta - 2\beta^{\frac{1}{2}}(\alpha + \delta)^{\frac{1}{2}}\right) \tag{22}$$

which is analogous to the one derived in Ref. 9.

For plasma to be unstable, the growth rate should be positive. This condition defines the critical value of the thermoelectric transport coefficient necessary for the instability onset:

$$\chi_0 > 2\mu_i T_0 \left(\varepsilon_{ion} + \left(\frac{5\mu_e}{\mu_i} \frac{\varepsilon_{ion} \nu_0}{\nu'_{ion}}\right)^{\frac{1}{2}}\right). \tag{23}$$

The expression on the right-hand side is positive. Therefore, the first observation from Eq. (23) is that the thermoelectric transport coefficient should also be positive [see Figure 1(c)]. In many cases, the Arrhenius form of the ionization frequency $\nu_{ion} = A \exp\left(-\frac{\varepsilon_{ion}}{T_e}\right)$ approximates well the ionization frequency obtained by solving the Boltzmann equation. Then, one can obtain that

$\frac{\varepsilon_{ion}\nu_0}{\nu'_{ion}} = T_0^2$ and Eq. (23) can be re-written as

$$\chi_0 > 2\mu_i T_0 \left(\varepsilon_{ion} + T_0 \left(\frac{5\mu_e}{\mu_i}\right)^{\frac{1}{2}}\right). \tag{24}$$

For the conditions of the present studies, $T_0 \sim 3 - 4$ eV [see Figure 2(b) and Figure 6(b)], $\mu_e \sim (1 - 2) \times 10^2$ m$^2$/eV-s (see Figure 1) and $\mu_i \sim 1.6$ m$^2$/eV-s. Then, Eq. (24) requires that $\chi_0 > 660$ eV-m$^2$/s. Figure 1(c) shows that this condition is satisfied for the case 0 and is not satisfied for the case 1. This explains the results discussed in Section III.B.

Based on the results presented in this section and Section III.B, we can conclude that the mechanism of discharge stratification at the given conditions is the following. If the thermoelectric transport coefficient exceeds the critical value defined by Eq. (23), then the thermoelectric flux starts dominating the flux of thermal conductivity. This leads to the energy flow towards the regions with the largest electron density and temperature from the regions with the smaller density and temperature. This leads to the increase of the electron temperature in these regions which, in turn, leads to the increase of the ionization frequency. The latter results in further increase of the electron temperature in these regions *etc*. The instability stabilizes when both fluxes are balanced by the electron heating by the electric field and the energy dissipation in collisions with neutrals.

### D. Influence of the non-linear dependence of the ionization rate coefficient on the electron density

This subsection briefly discusses the influence of the ionization rate coefficient on the electron density. Such dependence is obtained due to the influence of electron-electron collisions on the EEDF[25] and, consequently, on $k_{ion}$. In the present studies, we used a simple approximation proposed, for instance, in Ref. 26:

$$k_{ion}(n_e, T_e) = k_{ion}(T_e) \begin{cases} exp(n_e/n_c), & n_e < n_1, \\ exp(n_1/n_c), & n_e > n_1. \end{cases} \qquad (25)$$

Here, $k_{ion}(T_e)$ is the ionization rate coefficient obtained from Bolsig+, $n_c$ controls the rate of the non-linear dependence and $n_1$ defines the saturation value. In the present studies, we have used the values $n_c = 8 \times 10^{15}$ m$^{-3}$ and $n_1 = 10n_c$ as in Ref. 7. For the gas pressure of 0.5 Torr, the chosen value of $n_c$ implies the plasma ionization degree exceeding $10^{-6}$. Therefore, here we compare the case 1 (see Section II) obtained under the assumption of non-Maxwellian EEDF with the accounting for the Coulomb collisions and the Maxwellian EEDF (for the transport parameters, see Figure 1). In both cases, the ionization rate coefficient was determined by Eq. (25) with $k_{ion}(T_e)$ obtained for a given assumption on the EEDF.

Here, it is important to note that the striations formation driven by the Dufour effect is only obtained for small discharge currents when the electron-electron collisions are negligible. Hence, the thermoelectric transport coefficient is large, and the electron diffusion flux destabilizes the electron thermal diffusion flux in the electron energy balance equation.

Figure 8 shows the comparison between these two cases for the rf frequency of 13.56 MHz, background gas pressure of 0.5 Torr and three values of the rf voltage amplitude. Our simulation results have shown that for both cases, striations exist only in the narrow window of applied voltages (discharge currents). Namely, for the Maxwellian EEDF it is in the range $V_0 \approx 1200 - 2500$ V, while for the non-Maxwellian EEDF it is in the range $V_0 \approx 1400 - 2200$ V. In both cases, the discharge current is approximately in the range $100 - 200$ mA. This observation is in qualitative agreement with the experimental results reported in Ref. 7.

Figure 8(b) shows the comparison between the period-averaged ion densities obtained for $V_0 = 1600$ V. In both cases, the discharge stratification is obtained due to the non-linear dependence of the ionization rate coefficient on $n_e$. The electron diffusion term $\chi_e \frac{\partial n_e}{\partial x}$ is much smaller than the electron thermal diffusion flux $\kappa_e \frac{\partial T_e}{\partial x}$ which cannot destabilize the discharge (see discussion in Section III.C)  One can see the formation of seven striations for the non-Maxwellian EEDF and of five striations for the Maxwellian EEDF.

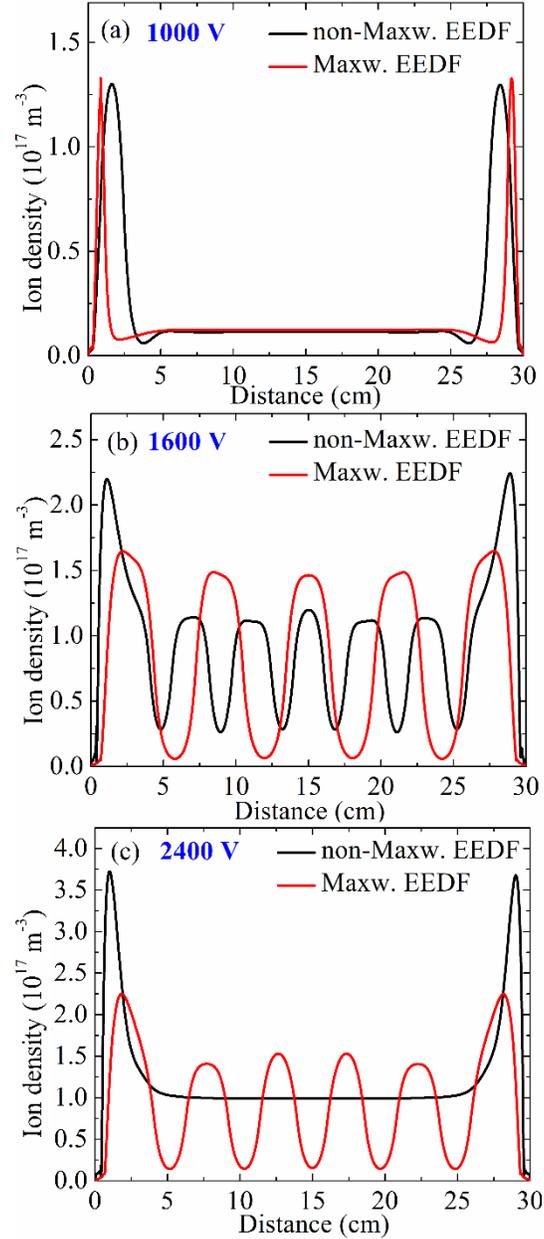

Figure 8. Period-averaged ion densities obtained for the rf frequency of 13.56 MHz for three values of the voltage amplitude: (a) 1000 V, (b) 1600 V and (c) 2400 V.

Figure 8(c) shows the comparison between the period-averaged ion densities obtained for $V_0 = 2400$ V. One can see that there is no discharge stratification for the non-Maxwellian EEDF. This is obtained because for this case the electron density within the positive column is $\sim 10^{16}$ m$^{-3}$ which exceeds $n_1$ in Eq. (25). Therefore, $k_{ion}$ saturates at $n_e > n_1$ and the ionization rate coefficient does not depend on the electron density. For the Maxwellian EEDF, the plasma density is below $n_1$ which results in the striations formation.

### IV.    Summary

The influence of Dufour effect on the striation of radio-frequency discharge was studied using the drift-diffusion fluid model for electrons and singly charged argon ions.

It is obtained that the striations formation is sensitive to the value of the thermoelectric heat transport coefficient in the range of electron temperatures $1 - 4$ eV typical for the positive column of radio-frequency discharge in long tubes. The critical value of the thermoelectric heat transfer coefficient was obtained using the linear stability analysis. The simulation results have shown that for the non-equilibrium electron energy distribution function the thermoelectric heat coefficient exceeds this critical value. Therefore, the standing striations originated at the sheaths edges and filled in the entire interelectrode space.

The accounting for the electron-electron collisions led to the thermalization of electrons which resulted in the decrease of the thermoelectric heat transfer coefficient. Consequently, its value dropped below the critical value for the instability onset. Therefore, the striations were not obtained.

Thus, the Dufour effect is responsible for the discharge stratification only for small discharge currents when the electron-electron collisions are negligible. For higher discharge currents, these collisions are non-negligible which results in the non-linear dependence of the ionization frequency on the electron density. This mechanism drives the discharge stratification at high discharge currents.

## AUTHOR DECLARATIONS

### Conflict of Interest

The authors have no conflicts to disclose.

## DATA AVAILABILITY

The data that support the finding of this study are available from the author upon reasonable request.

[1] P. S. Landa, N. A. Miskinova, and Yu. V. Ponomarev, Sov. Phys. Usp. **23,** 813 (1980).

[2] V. I. Kolobov, J. Phys. D: Appl. Phys. **39,** R487 (2006).

[3] Y. Liu, E. Schüngel, I. Korolov, Z. Donkó, Y. Wang, and J. Schulze, Phys. Rev. Lett. **116,** 255002 (2016).

[4] A. V. Nedospasov and N. V. Nenova, European Phys. Lett. **108,** 45001 (2014).

[5] H. Zhu, W. Yao, Z. Li, Plasma Process Polym. **17,** e1900271 (2020).

[6] V. Desangles, J.-L. Raimbault, A. Poye, P. Chabert, and N. Plihon, Phys. Rev. Lett. **123,** 265001 (2019).

[7] V. I. Kolobov, R. A. Arslanbekov, D. Levko and V. Godyak, J. Phys. D: Appl. Phys. **53,** 25LT01 (2020).

[8] R. R. Arslanbekov and V. I. Kolobov, Phys. Plasmas **26,** 104501 (2019).

[9] J. P. Boeuf, Phys. Plasmas **29,** 022105 (2022).

[10] V. I. Kolobov and R. R. Arslanbekov, Phys. Rev. E **106,** 065206 (2022).

[11] V. I. Kolobov, J. A. Guzman, and R. R. Arslanbekov, Plasma Sources Sci. Technol. **31,** 035020 (2022).

[12] D. Levko, Phys. Plasmas **28,** 013506 (2021).

[13] K. Bera, Sh. Rauf, J. Forster, and K. Collins, J. Appl. Phys. **129,** 053304 (2021).

[14] H. Urbankova, K. Rohlena, Czech. J. Phys. B **30,** 1227 (1980).

[15] N. A. Dyatko, I. V. Kochetov, and A. P. Napartovich, Plasma Physics Reports **37,** 528 (2011).

[16] N. A. Dyatko, I. V. Kochetov and A. P. Napartovich, Plasma Sources Sci. Technol. **23,** 043001 (2014).

[17] N. A. Humphrey and V. I. Kolobov, Plasma Sources Sci. Technol. **32,** 085017 (2023).

[18] G. J. M. Hagelaar, L. C. Pitchford, Plasma Sources Sci. Technol. **14,** 722 (2005).

[19] H. W. Ellis, R. Y. Pai, E. W. McDaniel, E. A. Mason, L. A. Viehland, Atomic Data and Nucl. Data Tables **17,** 177 (1976).

[20] D. L. Scharfetter and H. K. Gummel, IEEE Trans. Electron. Devices ED **16,** 64 (1969).

[21] G. J. M. Hagelaar, Modeling Methods for Low-Temperature Plasmas (Habilitation a Diriger des Recherches, 2009).

[22] L. S. García-Colín, A. L. García-Perciante, A. Sandoval-Villalbazo, Phys. Plasmas **14,** 012305 (2007).

[23] Yu. P. Raizer, Gas Discharge Physics (Intellect, Dolgoprudnii, 2009) (in Russian).

[24] V. I. Kolobov and V. A. Godyak, Plasma Sources Sci. Technol. **26,** 075013 (2017).

[25] Yu. Golubovskii, S. Valin, E. Pelyukhova and V. Nekuchaev, Plasma Sources Sci. Technol. **28,** 045015 (2019).

[26] R. R. Arslanbekov, V. I. Kolobov, Phys. Plasmas **26,** 104501 (2019).